\documentclass{article}

\usepackage{arxiv}

\usepackage[utf8]{inputenc} 
\usepackage[T1]{fontenc}    
\usepackage{hyperref}       
\usepackage{url}            
\usepackage{booktabs}       
\usepackage{amsfonts}       
\usepackage{nicefrac}       
\usepackage{microtype}      
\usepackage{lipsum}
\usepackage{graphicx}
\usepackage{multicol}
\usepackage{float}
\usepackage{enumitem}
\usepackage{siunitx}

\setlist[itemize]{leftmargin=*}

\title{Clusters of science and health related Twitter users become more isolated during the COVID-19 pandemic}

\author{
 Francesco Durazzi \thanks{F.D.\ and M.M.\ contributed equally to this work.}\\
  Department of Astronomy and Physics (DIFA)\\
  University of Bologna\\
  Bologna 40127, Italy\\
  \texttt{francesco.durazzi2@unibo.it} \\
   \And
 Martin M\"uller \footnotemark[1]\\
  Digital Epidemiology Lab\\
  Ecole polytechnique fédérale de Lausanne (EPFL)\\
  1202 Geneva, Switzerland\\
  \texttt{martin.muller@epfl.ch} \\
    \And
 Marcel Salathé \thanks{To whom correspondence may be addressed. Email: marcel.salathe@epfl.ch}\\
  Digital Epidemiology Lab\\
  Ecole polytechnique fédérale de Lausanne (EPFL)\\
  1202 Geneva, Switzerland\\
  \texttt{marcel.salathe@epfl.ch} \\
     \And
 Daniel Remondini \\
  Department of Astronomy and Physics (DIFA)\\
  University of Bologna\\
  Bologna 40127, Italy\\
  \texttt{daniel.remondini@unibo.it}  \\
}

\begin{document}
\maketitle

\begin{abstract}
COVID-19 represents the most severe global crisis to date whose public conversation can be studied in real time.
To do so, we use a data set of over 350 million tweets and retweets posted by over 26 million English speaking Twitter users from January 13 to June 7, 2020.
We characterize the retweet network to identify spontaneous clustering of users and the evolution of their interaction over time in relation to the pandemic's emergence.
We identify several stable clusters (super-communities), and are able to link them to international groups mainly involved in science and health topics, national elites, and political actors.
The science- and health-related super-community received disproportionate attention early on during the pandemic, and was leading the discussion at the time.
However, as the pandemic unfolded, the attention shifted towards both national elites and political actors, paralleled by the introduction of country-specific containment measures and the growing politicization of the debate.
Scientific super-community remained present in the discussion, but experienced less reach and became more isolated within the network.
Overall, the emerging network communities are characterized by an increased self-amplification and polarization.
This makes it generally harder for information from international health organizations or scientific authorities to directly reach a broad audience through Twitter for prolonged time.
These results may have implications for information dissemination along the unfolding of long-term events like epidemic diseases on a world-wide scale.
\end{abstract}

\keywords{COVID-19 \and scientific experts \and infodemic \and complex networks \and Twitter}
\newpage

\begin{multicols}{2}

\section{Introduction}
Twitter has been widely used as a tool for emergency response in previous crises and disasters and a large body of work has focused on optimizing communication in order to adjust or nudge human behaviour under such conditions~\cite{Chen2008,Li2010,Martinez-Rojas2018,Salathe2013}.
Research finds that during a time of crisis, when information is scarce and heavily sought after, social media enables the critical flow of information due to its collaborative nature~\cite{Graham2015,LeeHughes2009}.
However, the underlying network and community structure is likely to have a significant influence over which information users are exposed to~\cite{Sasahara2020,Conover2011,Newman2006}.
In particular, retweet interactions have been shown to reflect real-life community structure, such as political parties~\cite{Cherepnalkoski2016}, and to characterize the information sharing dynamics of different clusters of users~\cite{Bovet2019}. Likewise, communities of users sharing misleading news tend to be more connected and clustered~\cite{Pierri2020}.
Also tweet replies and quotes allow the reconstruction of a network structure, but the meaning and stance of these links is more ambiguous as compared to retweets, which typically imply agreement with or validation of the original message~\cite{Boyd2010}.

The structure of such information networks has been assessed in various disease outbreak scenarios, most prominently in the context of the Zika virus outbreak 2015--2016 in the US~\cite{Hagen2018}.
Among other findings the work suggests that key international experts acted as ``boundary spanners'', who were able to spread information quickly through the network.
Even though it has been observed that false information travels faster in social networks~\cite{Mendoza2010,Vicario2016,Vosoughi2018}, the work finds that Twitter users have made efforts to diffuse reliable information by such key experts.
Related work demonstrates how the association between the geographical location of users dictates topics of conversations, thereby matching the temporal and geographical spread of the Zika outbreak~\cite{Stefanidis2017}. 

Previously listed work allows for anecdotal evidence of crisis-induced behaviour on social media, but compared to the COVID-19 crisis the events were 1) shorter in time scale 2) more localized and 3) orders of magnitude smaller in terms of analyzed data.
As the virus spread across the world within only a few weeks, Twitter became the primary news source among expert groups and medical personnel.
Reliable sources, such as peer-reviewed literature or other officially vetted channels, were simply too slow to be useful during this time~\cite{Rosenberg2020}.
However, this initially positive role of Twitter and other social media has now been overshadowed by the spread of numerous conspiracy theories and other low-quality mis- and disinformation about the pandemic, peaking in what the WHO now considers an ``infodemic''~\cite{who2020,Cinelli2020}.
Preliminary work focused on this issue and used social network analysis in order to determine drivers of the conspiracy theory which claims a link between 5G and COVID-19~\cite{Ahmed2020}.
Although misinformation is a major concern, more recent work suggests that experts and authorities are being heard and have received disproportionate attention~\cite{Gligoric2020}, but that key specialists may experience low reachability~\cite{Mourad2020}.
Moreover, other work finds that key medical professionals and scientific experts may experience lower ``sustained amplification'', meaning that the attention given to this group has not been constant, and overall lower, compared to media outlets or key political figures~\cite{Gallagher}.
Our work attempts to clarify the somewhat ambiguous premises about the role of experts during the pandemic, focusing on the patterns of retweet interactions between users.  
Previous works on the topic classified Twitter users involved in the COVID-19 discussion with Natural Language Processing methods~\cite{Gligoric2020,Mourad2020} or user information extracted from external databases~\cite{Gallagher}, which allowed the labelling of only a fraction of the studied population. Our approach, instead, gathered all the users into network communities, which were a-posteriori associated to specific categories. In this way, with the help of a comprehensive dataset of more than 350 million tweets, we identify the key communities of English speaking Twitter users involved in the COVID-19 debate, starting from early January to June 2020.
We provide a detailed analysis of the evolution of this massive communication network, in particular regarding the interaction dynamics within and between the communities, showing how the attention toward these super-communities follows different trajectories over time.

\section{Results}
\subsection{Network community identification}
In the giant component of the directed network (see Methods), the out-degree distribution (i.e.\ number of retweets received by each user) follows a very skewed distribution, that  can be fitted by a power law with exponential cut-off (see Supplementary Fig. S1). 
Users with an out-degree higher than 1500 represent the 0.1\% of the users in the network, but their tweets have been retweeted >200M times (77.0\% of all retweets).
The community detection algorithm reveals thousands of communities, spanning from millions down to duplets, with size decreasing very sharply (see Supplementary Fig. S2).
By combining multiple repetitions of the clustering algorithm (see Methods), we identify 15 communities (labeled with letters from A to O, in decreasing order of size) with more than $10^5$ users, encompassing 97.9\% of all users in the giant component. 

\end{multicols}
\begin{figure}
\centering
\includegraphics[width=\textwidth]{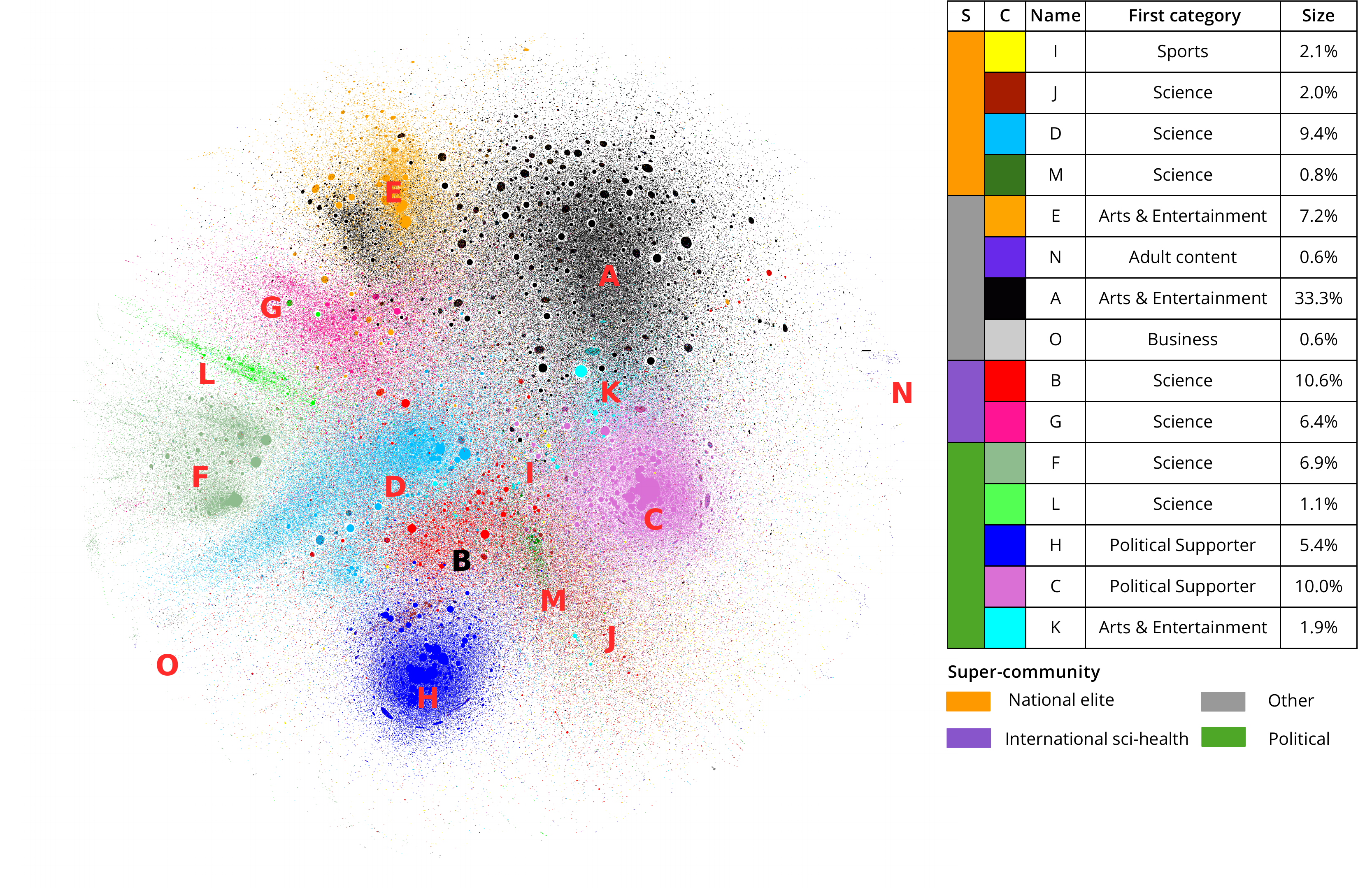}
\caption{
Retweet network of a randomly sampled connected component of 1M users, colored by community.
Node size is proportional to node out-degree.
In the table, the column ``S'' designates the color of the super-communities used throughout this work, ``C''  lists the community color in the network layout and ``Name'' the respective community name.
The ``Dominant category'' column specifies the most abundant user category in the community (excluding ``Other'').
``Size'' denotes the ratio of users in the community with respect to the total number of users in the network. 
}
\label{fig:1}
\end{figure}
\begin{multicols}{2}

Figure~\ref{fig:1} shows the network of 1M users (corresponding to 4.3\% of all nodes) randomly sampled but forming a connected component.
Even if the layout algorithm has no information about the communities we identified, it recovers a similar stratification of the nodes.
We observe that communities A and B fill a great portion of the network, without being as densely connected as the other communities.
Some communities (C, H, F, L) are more peripheral and exhibit a neat modular structure, which implies a high degree of internal connectivity. 

\subsection{Characterization of communities}
In order to characterize the 15 largest communities, we 1) infer the presumed location of the users and 2) use the results of a Machine Learning classifier, which was trained to predict the user category based on a user's description (see Methods).
Self-reported location information was available for 54\% of nodes in the network.
Since the diversity in nationality of the communities is of interest, we calculate the entropy of the distribution of countries for each community (similarly to the alpha diversity index in ecological communities): a high entropy indicates an almost uniform distribution of users across all countries, whereas a low entropy implies an uneven distribution, possibly skewed toward a specific country.
Some communities are strongly associated with user location, in particular communities D (UK), E (Philippines and Southeast Asia),  F (India), G (African countries), J (Canada), L (Pakistan) and M (Australia).
The US represents more than a half of the users from communities C, H, I, and K, while communities A, B, N and O are more heterogeneous in terms of location, containing a large fraction of tweets located outside of the US. 

In some communities, the most retweeted users have a strong association to political and cultural topics.
In order to better understand these mixed communities, we investigate the predicted user categories for each community (see Methods, Table S1 and Supplementary Fig. S3).
Out of the 13 categories, the category ``Other'' is the majority category for all considered communities, being assigned to 78\% of all users and having received 22\% of all retweets.
Science, being the second largest category (at 4.7\% of all users), was represented at 9\% or higher in communities B, D, G, J and M.
Other communities, such as H or C, have a higher fraction of users who identify with a political movement.

\begin{figure}[H]
\centering
\includegraphics[width=\linewidth]{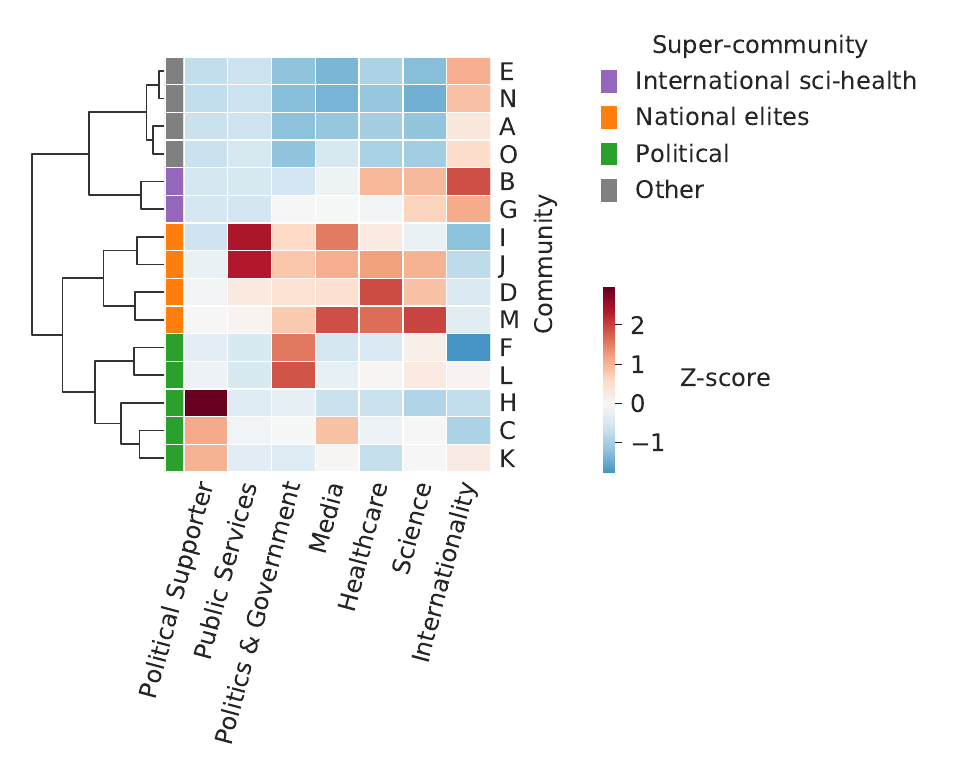}
\caption{
Clustering of communities into super-communities.
The heatmap shows the Z-scores (i.e.\ standardized values) for seven chosen features.
The four super-communities denote the emerging clusters.
}
\label{fig:2}
\end{figure}

In this study, we are particularly interested in the specific user categories ``Science'', ``Healthcare'', ``Media'', ``Politics \& Government'', ``Public Services'', and ``Political Supporter''.
The percentage of users in these six categories and the entropy measure for internationality were used for hierarchical clustering at the community level (see Figure~\ref{fig:2} and Material and methods section).
With the help of the resulting dendrogram, we recognize four super-communities, which we name based on the categories over-represented in each super-community (Z-score > 0):
\begin{itemize}
    \item International sci-health: communities exhibiting an increased number of users of category ``Science'' or ``Healthcare'' and a high international diversity.
    \item National elite: communities exhibiting an increased number of users belonging to official categories (i.e. ``Healthcare'', ``Media'', ``Politics \& Government'', and ``Public Services'') and a low international diversity.
    \item Political: communities exhibiting an increased number of users associated with political activism only (category ``Political Supporter'') or involved in politics (``Politics \& Government'').
    \item Other: communities which are not linked to any of the categories of interest, because no category has a Z-score > 0 except ``Internationality''. This super-community includes artists, entrepreneurs, and non-governmental activists.
\end{itemize}

The naming of super-communities is based on the over-represented categories listed above, and further validated by the manual inspection of the top users in the respective communities.
In particular, only two communities (B and G) are assigned to ``International sci-health'', showing a clearly distinguishable pattern.
These communities consist of a large fraction of users who are presumably working in scientific or health-related fields or who frequently retweet top medical and scientific expert.
Community B's top users include well-known news agencies, as well as the WHO, making this community led by official media and scientific information spreaders.
Community G is similar, but it has more users from African countries, in particular Nigeria and South Africa.
National elites, i.e.\ communities I, J, D and M, are communities with a high proportion of official categories not only Science and Healthcare related, but linked to specific countries.
Among the political super-community, communities F and L have the highest proportions of users and institutions involved in politics, while communities C, H and K are driven by US-specific political debates.
Upon visual inspection of a sample of accounts, it emerges that community C and K are more often associated with the Democratic party, and H with users from the Republican party.
All other communities, including community A (comprising 32\% of all users), show characteristics which were not deemed relevant for this analysis.

\begin{figure}[H]
\centering
\includegraphics[width=\linewidth]{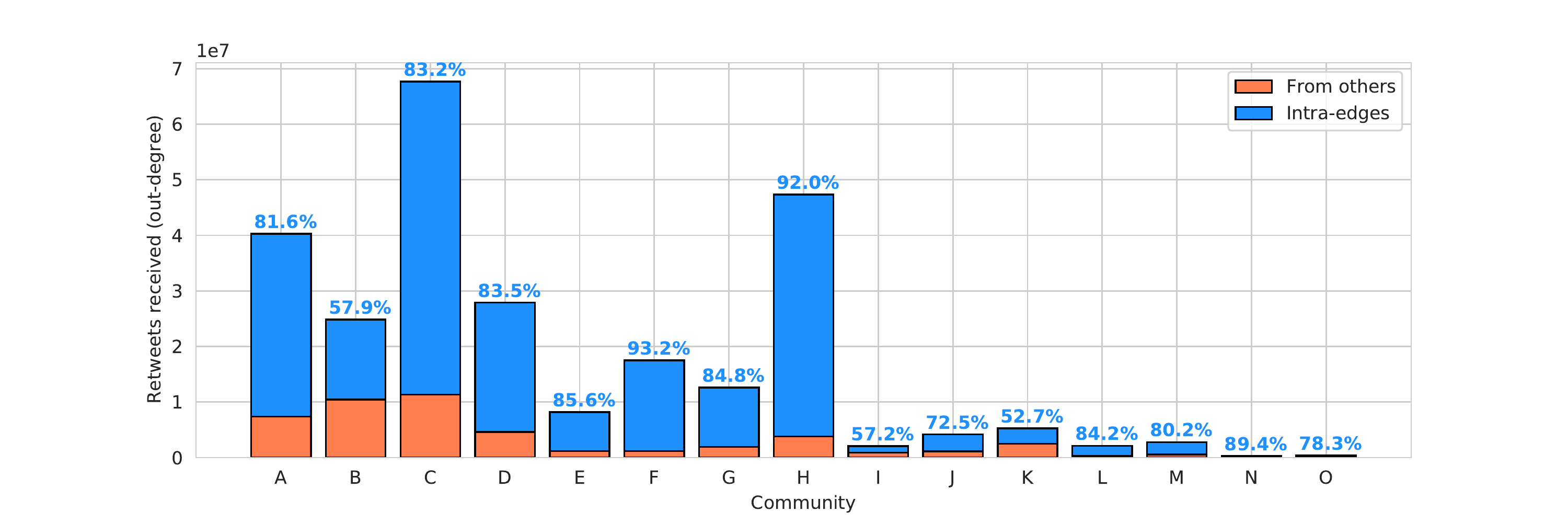}
\caption{
 Internal and external components of the attention towards each community. Each bar represents the number of total retweets received by users in each community, divided into retweets from the same (blue, with the percentage represented on top of each bar) and from other (orange) communities. 
}
\label{fig:attention}
\end{figure}

We quantify the attention received by the communities through the number of retweets their users received (Figure~\ref{fig:attention}). This measure can be decomposed into an external and an internal component, depending on whether the retweets are received from outside or inside the community, as further investigated in the next section. Community C and H, assigned to the political super-community, are the most retweeted communities, though not the largest. The attention received by these two communities is mainly internal (more than $80\%$). Community B, assigned to the sci-health super-community, is the second largest one in terms of number of users but received low overall attention ($5^{th}$ most retweeted in total). Nevertheless, it has a high level of reach, ranking $2^{nd}$ with respect to the external attention component.

\subsection{Network dynamics}
\begin{figure}
\centering
\includegraphics[width=\textwidth]{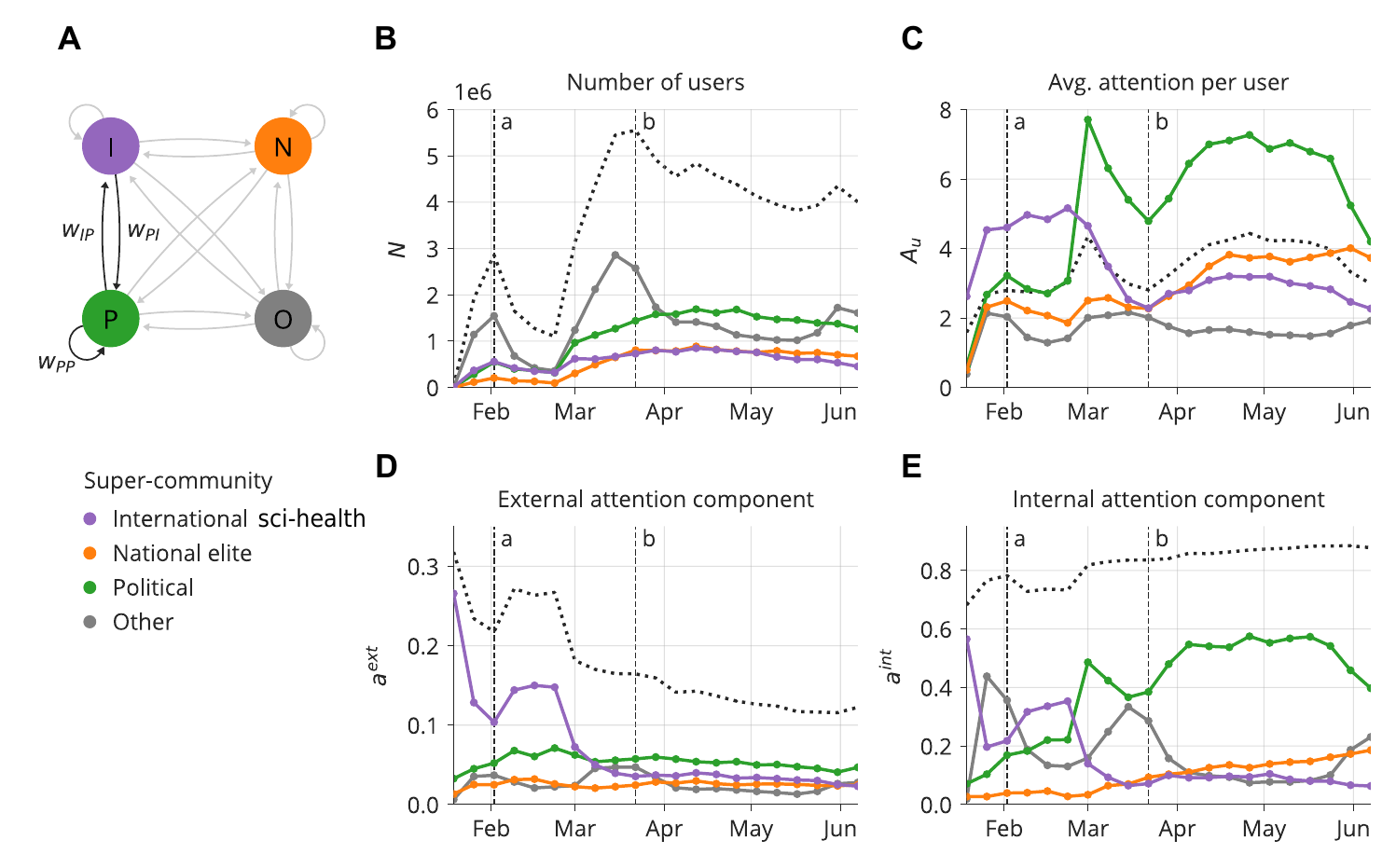}
\caption{
 Evolution of weekly aggregated networks by super-community, with dotted lines corresponding to the statistics across all users.
 \textbf{(a)} Diagram representing the networks collapsed to the super-community level.
 Edge direction represents the flow of information via retweets, i.e.\ from retweeted to retweeting super-community.
 \textbf{(b)} Size of the super-communities in terms of number of users.
 \textbf{(c)} Average attention per user.
 \textbf{(d)} External component of the attention toward super-communities.
 \textbf{(e)} Internal component of the attention toward super-communities.
 Indicated as $a$ and $b$ are the first and second peak in terms of network size, as shown in Figure~\ref{fig:3}b.
 }
\label{fig:3}
\end{figure}

In order to trace the fluctuation of the retweet dynamics between users throughout the pandemic, we reconstructed several networks by splitting the Twitter data into non-overlapping windows of 1 week, with the purpose of having a reasonable numbers of time windows with enough statistics within the observation period.
We collapsed the 1-week retweet networks into networks with four nodes, aggregating the links between users previously assigned to the four super-communities (see Figure~\ref{fig:3}a) and thereby compute the so-called mixing matrix~\cite{Newman2010}.
In these networks, we draw a link from super-community \textit{i} to \textit{j}, with a weight $w_{ij}$ equal to the number of times super-community \textit{j} retweeted \textit{i} in a given week. 

We assign each node a size attribute $N_i$, representing the number of users of super-community $i$ retweeting or being retweeted in the given week.
Figure~\ref{fig:3}b shows the temporal evolution of this attribute.
Two principal observations can be made: first, the value of $N$ varies over time in correspondence with the phases of the epidemic.
We identify an initial peak in total number of users in the beginning of February (peak $a$), and a second one at the end of March (peak $b$).
These two peaks presumably correspond to the first diffusion of news about COVID-19 in China and, later, to worldwide diffusion of the pandemic (see Supplementary Fig. S4).
In the time between peak $a$ and $b$ the number of users in the COVID-19 debate has doubled, followed by a slow decay between April and June.
Second, most of the change in the number of users stems from the ``Other'' super-community.
This implies that the three super-communities of interest remained present throughout the entire observation period at a relatively static level.

In Figure~\ref{fig:3}c, we show the average attention per user of super-community $i$, defined as:
\begin{equation}
    A^i_u=\frac{\sum_j w_{ij}}{N_i}
\end{equation}
i.e.\ the weighted out-degree of super-community $i$, normalized by its size.
The international sci-health super-community faces an increase in average attention per user in January and stabilizes at a higher level compared to other super-communities until the beginning of March.
After a narrow peak in March, the political super-community plateaus in April at roughly three times the attention level of national elites and international sci-health.

We then split the total attention, i.e.\ the sum of all weighted edges $W$, for each super-community $i$ into an internal and external component for every weekly network:
\begin{equation}
    a_i^{ext}=\frac{\sum_{j\neq i} w_{ij}}{W}
\end{equation}
\begin{equation}
    a_i^{int}=\frac{w_{ii}}{W}
\end{equation}
\begin{equation}
    a_i^{ext}+a_i^{int}=1
\end{equation}
The external component $a_i^{ext}$ represents the attention given to super-community $i$ from the other super-communities, while the internal component $a_i^{int}$ quantifies self-amplification.
Figure~\ref{fig:3}d shows that the external attention component is decreasing overall, indicating a decrease in attention between super-communities.
This is particularly true for the international sci-health super-community, which received broad attention in the very beginning of the pandemic, peaking again in mid-February, and then decaying in a monotonic way until the end of our sampling, while the pandemic was spreading to all the countries worldwide.  

\end{multicols}
\begin{multicols}{2}

Figure~\ref{fig:3}e shows that the internal attention component is increasing overall, highlighting an increased self-amplification within the network.
We observe that this increase is mostly driven by the political super-community and to a lesser degree by the national elites, suggesting also an increase of political polarization within the retweets network, at the poles of which the antagonistic political communities lie.
The international sci-health super-community, on the other hand, decreased internal sharing of content after March.
Overall, we note that the dynamics partially mirror Figure~\ref{fig:3}c, since the internal attention component makes up most of the total attention given to the super-communities.

Considering the average number of tweets per user within each super-community (defined as "activity", see Supplementary Figure S5), we observe that while the activity increased over time for the political and national elites super-communities, there was a reduction for the international sci-health super-community.

To further characterize the evolution of the attention patterns, we performed an analysis of the network community structure dividing the observation period into monthly windows. Moreover, since it is known from literature that critical events may affect the sentiment of the discussion~\cite{Garcia2019} and therefore impact the community structure~\cite{Mitrovic2010}, we quantified on a weekly basis the variation of sentiment ("subjectivity" and "polarity") in tweets' content over time by using a deep learning and a a lexicon-based algorithm (see Material and methods section). These analyses (detailed in the Supplementary Text 1.3) showed a stable network community structure, and no significant trend in sentiment was observed over time. Regarding this last statement, we remark that the null result may be due to sentiment analysis algorithms not specifically tuned to COVID-related topics. This limitation, in particular, has already been observed for lexicon-based algorithms~\cite{Zimbra2018}.

\subsection{Sustained attention towards top users}

So far, in our analysis the observed dynamics of super-communities is a result of the average behavior across all the users in the Twittersphere, while in reality most of the dynamics are driven by a relatively small set of users who receive disproportionate attention.
Further, we have focused on the number of retweets (node out-degree) as a canonical measure of attention, but the retweet count of a single viral tweet might exaggerate the user's real impact in the overall debate.

In this section, we address these caveats.
Here, we only consider the top 1000 users for each super-community in terms of retweets received, i.e.\ 4000 users receiving 55.0\% of all retweets.
Furthermore, we adopt an alternative measure to characterize the attention given to users, namely a retweet $h$-index, as previously introduced by ~\cite{Grcar2017} and used by ~\cite{Gallagher} in the context of COVID-19 posts on Twitter.
Originally proposed in the context of academic citations~\cite{Hirsch2005}, the $h$-index in this case reflects that the user has received at least $h$ retweets on $h$ of their original tweets.

\begin{figure}
\centering
\includegraphics[width=.8\textwidth]{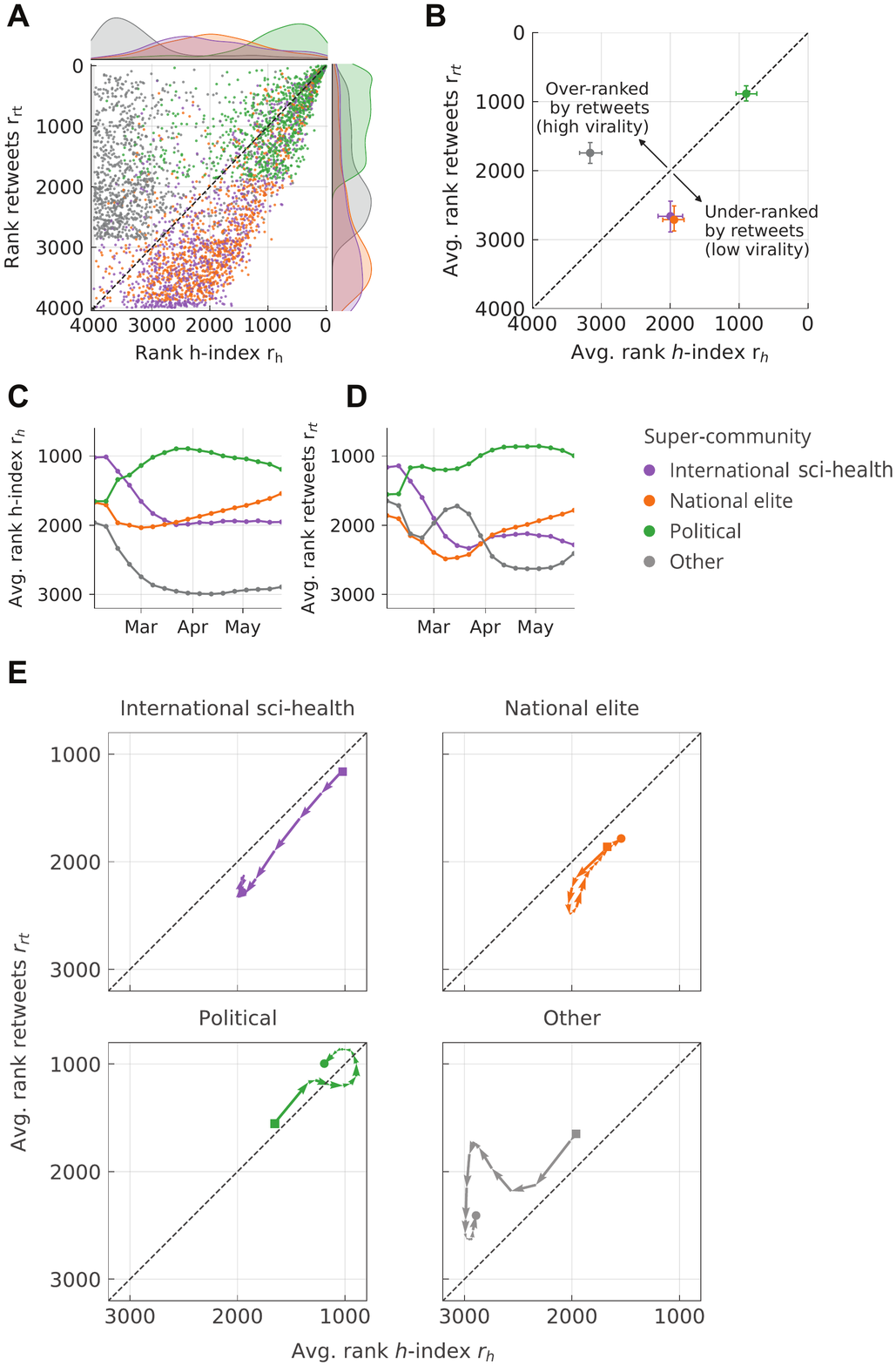}
\caption{
Static and temporal communication patterns of the top 1000 users of each super-community.
\textbf{(a)} Comparison between the rank in terms of $h$-index and retweets by user, as well as respective marginal distributions.
\textbf{(b)} The average rank by super-community with bootstrapped 95\% confidence intervals.
\textbf{(c)} Weekly $h$-index rank computed within a rolling time window of 1 month. 
\textbf{(d)} Weekly rank in retweets computed within a rolling time window of 1 month.
\textbf{(e)} Vector plot of retweet and $h$-index ranks by super-community.
Each arrow denotes the change in ranks within one week. The first and last week are marked with a square and circle, respectively.
}
\label{fig:4}
\end{figure}

Figure~\ref{fig:4}a and~\ref{fig:4}b compare the rank in terms of retweets received $r_{rt}$ and $h$-index $r_h$ both on the user and the super-community level, respectively.
A user placed in the top-left in the figure plane suggests that few of their tweets received punctual attention at high virality, whereas the bottom-right suggests sustained or long-lasting attention at low virality.

\end{multicols}
\begin{multicols}{2}

Generally, users belonging to the political super-community are ranked highest both in terms of retweets and $h$-index, receiving most of the attention both on a punctual and an extended time scale.
National elites and international sci-health super-communities behave very similarly: they rank medium to high in terms of $h$-index, but low in the number of retweets received (low virality).
Lastly, the ``Other'' category ranks generally lowest in $h$-index and intermediate in retweets, thus is characterized by a higher virality in terms of attention.

In order to understand the temporal dimension of these results, we formulate the previous metrics in a time-dependent fashion on the same set of top users.
We consider all tweets and retweets posted within a rolling time window of 1 month width and a 1 week step.
We then compute the rank by $h$-index (Figure~\ref{fig:4}c) and retweets (Figure~\ref{fig:4}d) averaged by super-community.
Additionally, we show the resulting data as a vector plot (Figure~\ref{fig:4}e).
We find that international sci-health users scored very highly in both metrics initially and then experienced a drop in ranks.
Similarly, national elites faced an initial decline in both metrics but then increased in ranks above the level of international sci-health after April.
The temporal view of this data therefore adds nuance to the picture obtained in Figure~\ref{fig:4}b.

We conclude that although national elites and the international sci-health super-communities share a considerable overlap in the static view (cf.\ Figure~\ref{fig:4}b), they reveal distinct temporal dynamics (cf.\ Figure~\ref{fig:4}c--e).
This result reflects how the international sci-health super-community faced a decline in attention, while the discussion has been moving onto more local grounds, with the political and national debate gaining momentum. 

\section{Discussion}
In this work we reconstruct and describe the complex network structure of the user retweets in relation to the COVID-19 pandemic, and study the evolution of the interactions between and the attention towards its communities. We also make use of the network structure to characterize the role of international communities involved in science and health related topics in the English-speaking Twittersphere.\\
Using a community detection algorithm, we identify 15 user communities, exhibiting a stable structure throughout the observation period.
While previous studies rely on user descriptions to cluster them (providing a proper labelling of only a fraction of the users), we firstly identify communities based on the whole retweet link structure, then we find the prevalent user categories within each community, thus allowing the classification of all users in the available dataset.
We are able to group these communities into four main super-communities related to the prevalent user categories and the degree of internationality (namely international sci-health, national elite, political, and other) and assess their interaction patterns over time, focusing on the flow of intra- and inter- community retweet volume.

In the Twitter landscape of COVID-19 we identify a single major group with interests in scientific and health topics and with a highly international distribution of users, and multiple country-specific communities which appear to engage more in the respective national debates.
Additionally, we find several large country-specific communities which are mostly characterized by political activism, thus highlighting the substantial politicization in the discussion surrounding the pandemic.

Our results emphasize the role of the international sci-health super-community in the beginning of the, at the time, largely unknown pandemic.
This super-community received disproportionate attention and had broad reach across many cultural and political communities, as reflected by the high total volume of retweets received, in particular from users that we characterized as belonging to non-scientific communities.
As demonstrated by the high internal attention component, the users of this super-community also shared content frequently among themselves, possibly in an attempt to rapidly share scientific insights about the novel virus.

In a second critical moment in March, the number of COVID-19 cases exploded in almost all parts of the world, leading to massive media attention, which is well reflected in the increase in community sizes, as well as in the total number of tweets.
It is noteworthy that this added attention has not been allocated to international sci-health experts anymore but rather to local political leaders, reflecting $1)$ the loss of global influence of the former in the evolution of the debate, and $2)$ the fact that users were redirecting their attention towards their local authorities and referents.
This might be due to the fact that COVID-19 pandemic was no more an external phenomenon arousing a general interest, but it was starting to directly impact people's lives.
In conjunction to this shift in attention, the international sci-health super-community decreased its activity (original tweets per user), while for national elites and political super-communities it increased.

As the pandemic unfolded in April, and while many English speaking countries underwent a strict lockdown, we observe a growing politicization of the debate, reflected by the fact that content by the political communities is now shared most.
Meanwhile, the analysis reveals the picture of a less retweeted international sci-health super-community.
Furthermore, compared to January and February, and in contrast to all other groups, this super-community also reduced its activity and the interactions among themselves, even though their size remained constant.

The analysis of the sustained amplification patterns (measured by the retweet \textit{h}-index) of highly retweeted users extends our previous analysis performed at a super-community level: compared to political leaders, top users in the international sci-health super-community received intermediate levels of sustained attention but their content lacked virality.
Additionally, the results show the importance of the national elites in influencing the discussion after the end of March: national elites’ top users show a positive trend in both punctual and sustained attention received, to the point of surpassing the international sci-health super-community.

Our work, characterizing the evolution over time of these observables, provides an explanation for the discrepancies found in recent literature on the role of scientific information spreaders in the first months of the COVID-19 pandemic. By characterizing the retweet network structure dynamics, in parallel to the worldwide spreading of the pandemic, we are able to confirm that the scientific sci-health super-community was mostly heard early on in the pandemic, as found in~\cite{Gligoric2020} which utilized the same algorithm for user classification.
This also fits in well with previous literature, which confirms that Twitter users amplify information from trusted sources in crisis situations~\cite{Reuter2018,Wagner2018,Shin2016}.
Further, we find that the attention towards sci-health users has declined over time, whereas other authors~\cite{Mourad2020} indicated a low reach of doctors and experts, without acknowledging the leading role in the first months of international communities interested in science and healthcare.
For the same reason, we can only partially confirm previous work~\cite{Gallagher} that suggested low sustained attention to top medical experts.\\
We do not know the exact overlap between our sci-health super-community and the expert groups defined in the previous papers, thus it is hard to make a direct comparison. Nevertheless our study identified the communities gathered around the top sci-health users and added a temporal dimension to the analysis, showing that the sci-health super-community ranked highest both in sustained and punctual attention in the very beginning of the pandemic and only later became increasingly isolated in terms of attention.
Our results underline the different role of sci-health over time, seen as possible boundary spanners only during the early phase of the pandemic. 
We observed that the sentiment of tweet content remained quite stable over time. This might suggest that the cause of this drop in attention is not due to a different emotional response or a change in perception of the messages. It's possible that the algorithms we used for sentiment analysis, which were not specifically trained on COVID related text, may have not been sensitive enough. Overall, the international sci-health super-community lost some weight in the discussion when the epidemic issue became more impactful on a local scale and politics came into play, attracting more attention, and possibly because it reduced the tweeting activity compared to the first months. Anyway, the emerging political and national authorities did not have the same degree of inter-community reach, making the discussion more compartmentalized.

\section{Conclusion}

Our work leads to two main conclusions: 1) under the unique circumstance of an emerging virus causing a global pandemic, the Twitter platform allowed thousands of international users with experience and interest in sci-helath topics (including leading experts such as the WHO) to quickly and efficiently exchange information, while also being amplified by non-scientific communities. This effect was localized only in the early phase of the pandemic, before its spread worldwide, since 2) as the pandemic developed, Twitter users directed more of their attention towards the national debates, overall leading to less interacting communities with respect to the initial phase.

Our study is based on a comparatively large dataset encompassing a total of 354M tweets by 26M users, which can be considered as comprehensive (see Methods).
However, a limitation of this work is that a substantial part of the network consists of the super-community labelled as ``Other'', encompassing around 50\% of all users.
It is difficult to make general statements about the true nature of this group, as it includes a very diverse set of users, with most of them reporting unspecific profile information.
Our work is based on the self-reported status of users, thus future research could be required to properly understand their true nature.

As a final consideration, we can hypothesize  that, as long as the COVID pandemic was a novelty, Twitter users from all the super-communities tended to give more attention to the messages from the sci-health super-community because it was initially considered as the most authoritative reference in this topic. Our network analysis highlights that in this first phase the amount of internal debate within political and national elite communities was reduced, possibly due to the fact that the pandemic was not impacting all local levels directly. After a few months from its emergence, the epidemic started to create social and health problems in many countries, prompting Twitter users to turn their attention to politics and national issues.
It also is noteworthy that the international sci-health super-community itself reduced the level of self-amplification and its tweeting activity. We verified that this attention shift can not be clearly explained by a change in the sentiment of the tweet content, since the algorithms we used provided a constant output over the observed time period.

Our results suggest that an exclusively scientific discussion risks losing audience when the health crisis starts to heavily impact society and to feed country-specific debates, thus scientists and health institutions should maintain a regular tweeting activity and reshape their valuable content in order to involve themselves usefully in local discussions, targeting and merging with the stable national Twitter communities.

\section{Materials and methods}
\subsection{Data collection}
\label{subsec:data_collection}
Twitter data was collected through the Twitter API, specifically through the filter streaming endpoint, using the Crowdbreaks platform~\cite{Muller2019}.
The data used in this work consists of a total of \num{353 993 900} tweets (thereof \num{267 026 740} retweets) posted by \num{26 262 332} users in a 146 day observation period, i.e.\ from January 13 to June 7, 2020.
These tweets have been identified by Twitter to be in English language and match one or more of the keywords ``wuhan'', ``ncov'', ``coronavirus'', ``covid'' and ``sars-cov-2''.
Note that keywords have changed over time, as the new names for virus were introduced (for details refer to Supplementary Text 1.1 and Supplementary Table S2).
The data is complete with respect to these keywords, except during a period between mid-March to mid-April when volume exceeded the 1\% threshold imposed by Twitter and was subsampled by an (unknown) degree.  

\subsection{User categorization}
\label{subsec:user_categorization}
In order to be able to interpret the identified network communities, accounts were categorized by occupational role and account type.
This categorization was conducted using a Machine Learning classifier which was trained to determine the category of an account solely based on the user description (user bio). The category assigned to each user is based on self-reported information, though it reports a self-declared expertise or interest in a particular category more than a verified expertise in it.
The classifiers were first published in the context of related work on the attention given to experts during COVID-19~\cite{Gligoric2020}.
In this work, we use the published English language BERT model (\texttt{bert-english-pt}) in order to determine the category of each account in our dataset.
In the aforementioned study~\cite{Gligoric2020}, the categories have been determined in an iterative coding process to best categorize users into 13 categories, which are: Adult content, Arts \& Entertainment, Business, Healthcare, Media, Non-governmental organization (NGO), Political Supporter, Government \& Politics, Public Services, Religion, Science, Sports, and Other.
For further details on the coding process or the training of the machine learning classifiers, please refer to the referenced study~\cite{Gligoric2020}.

\subsection{Geo-localization}
Tweet objects contain both structured and unstructured forms of geographical information.
In this work, we employed a procedure to geo-localize tweets on the country level using the Python library local-geocode (\url{https://github.com/mar-muel/local-geocode}), please refer to Supplementary Text 1.2 for detailed explanations).
A user's country location was determined from the majority of the user's geo-localized tweets.
Geolocation could be inferred for 75\% of tweets belonging to 54\% of the considered users.

\subsection{Network analysis}
\label{subsec:network_analysis}
We study the full directed retweet network, consisting of all retweets collected during the entire 147 day observation period (267M retweets).
The nodes of the network represent users who have at least once retweeted or have been retweeted by another user.
An edge was established from user A to user B if B retweeted A at least once during the whole period of data collection.
Therefore, the edge direction indicates the flow of information: the higher the out-degree, the more the user has been retweeted.
We assigned a weight to this edge equal to the number of times user B has retweeted user A.
The reconstructed (weighted and directed) network has 22.9M nodes and 177M unique edges.
In order to study the communities in this network, we consider only the largest connected component of the network, consisting of 22.5M nodes and 176M unique edges.
The discarded components consisted of small star-like communities (size range 1-280, median:2), making up 1.57\% of the nodes and 0.13\% of the edges in the original network. 
Since Twitter, as many other social networks, is characterized by a high degree of homophily, defined as the tendency of similar individuals to form ties with each other~\cite{Colleoni2014,Williams2015}.
We made use of Louvain algorithm for community detection~\cite{Blondel2008}, which maximizes the network modularity. 
We symmetrized the network since we were not interested in the direction of retweet flux, but rather we constructed communities on the total volume of tweet exchange between users that defines the tightness of their bond. We used the Python package Networkit~\cite{Staudt2016} for its efficiency in handling large networks, considering the default for the resolution parameter $\gamma=1$.
Due to the stochasticity of the clustering algorithm, we ran 50 trials with different random initialization and assigned each node to the community it was most frequently associated with.
The identification of largest communities was found to be stable both among repeated runs of the algorithm (Supplementary Fig. S6).
In the main text, we show the results for the network obtained from the whole time-series of retweets. In Supplementary Text 1.3 we show that networks reconstructed from monthly time-windows are very stable over-time both at community and super-community level.
Thus, the results of the community detection can be considered as fairly robust.
The coordinates of the network layout in Figure~\ref{fig:1} were processed by Gephi software~\cite{Bastian2009}, using the ForceAtlas2 algorithm~\cite{Jacomy2014} with gravity set to 0.05 with the ``stronger gravity'' option enabled. The out-degree distribution of the network (Supplementary Fig. S1) was analyzed through the package powerlaw~\cite{Alstott2014}.

\subsection{Sentiment analysis}
We performed sentiment analysis through the Python's library TextBlob, which operates a lexicon based analysis of the text to compute the subjectivity and the polarity of a message~\cite{loria2018}. This algorithm assign to each original tweet 1) a polarity score, corresponding to a sentiment which spans from negative (-1) to positive (+1), and 2) a subjectivity score, where a value of 1 generally refer to personal opinion, emotion or judgment whereas a value of 0 refers to objective and factual information. We then classified each retweet in the dataset as Negative, Neutral or Positive depending on whether the shared tweet had polarity less than -1/3, between -1/3 and +1/3 or higher than 1/3 respectively. For comparison, we also applied a roBERTa deep learning model for sentiment classification into negative, neutral and positive sentiment~\cite{barbieri2020}.

\subsection{Super-communities definition}
We mapped each community detected on the retweet network to a vector concatenating the relative abundance of each user category within the community and the internationality measure. We standardized the features (Z-scores) and built an agglomerative hierarchical dendrogram with the Ward's method for variance minimization. We identified the knee point of the average inter-clusters distance at 3 clusters, through the maximum of the 2\textsuperscript{nd} derivative (see Supplementary Fig. S7). The emerging clusters were [E,N,A,O,B,G], [I,J,D,M], [F,L,H,C,K]. We decided to split the first cluster into [E,N,A,O] and [B,G] because of the overabundance of "Science" users in the latter 2 communities (Z-score > 0), which is a relevant category for the topic of interest. Furthermore, [E, N, A, O] are not over-represented by any of the categories of interest, but rather populated by users associated to  Sports, Adult content, Business, Arts \& Entertainment (see Supplementary Fig. S3).

\paragraph{Data availability.}
The datasets and the code generated and/or analysed during the current stury are available through the public GitHub repository \url{https://github.com/FraDurazzi/twitter-network-covid19}.
The full Twitter dataset used in this work is available on Zenodo~\cite{muller_martin_2020_4267033}: \url{https://doi.org/10.5281/zenodo.4267033}. More information is available upon reasonable request from the authors.

\paragraph{Author contributions.}
M.M.\ collected the data.
F.D.\ and M.M.\ analyzed the data.
F.D., M.M., D.R.\ and M.S.\ designed the study and wrote the paper.

\paragraph{Competing interests.}
The authors declare no competing interests.

\paragraph{Acknowledgments.} 
We thank Marion Koopmans for her valuable comments and ideas that helped to design the study.
This project was funded through the European Union's Horizon 2020 research and innovation programme under grant agreement No.\ 874735 ``Versatile emerging infectious disease observatory - forecasting, nowcasting and tracking in a changing world (VEO)''.

\bibliographystyle{unsrt}  

\bibliography{main}

\end{multicols}

\end{document}